\documentclass[usegraphicx,usenatbib]{mn2e} 
\usepackage{epsfig}
\usepackage{amsmath, amssymb}
\usepackage{natbib}

\newif\ifAMStwofonts

\newcommand{\OIII}{{\rm[OIII]}}
\defcitealias{Merloni2003}{M03}
\defcitealias{Heckman2005}{H05}

\title[Testing Black Hole Jet Scaling Relations]
      {Testing Black Hole Jet Scaling Relations in Low Luminosity AGN}

\author[F.~De~Gasperin et~al.]{F. de Gasperin$^{1,2}$,  A. Merloni$^{3,4}$, P. Sell$^{5}$,
P. Best$^{6}$, S. Heinz$^{5}$ and G. Kauffmann$^{1}$\\
$^{1}$Max-Planck-Institut f\"ur Astrophysik, Karl-Schwarzschild-Str. 1, D-85741, Garching, Germany\\
$^{2}$Exzellenzcluster Universe, Boltzmann-Str. 2, D-85748, Garching, Germany\\
$^{3}$Max-Planck-Institut f\"ur Extraterrestrische Physik, Giessenbachstr., D-85741, Garching, Germany\\
$^{4}$Technische Universit\"at M\"unchen, James-Franck-Str., Garching, Germany\\
$^{5}$University of Wisconsin-Madison, 475 N. Charter St., 53706 Madison, WI, USA\\
$^{6}$Institute for Astronomy, University of Edinburgh, Blackford Hill, EH9 3HJ, Edinburgh, UK}

\begin{document}

\date{}
\pagerange{\pageref{firstpage}--\pageref{lastpage}} \pubyear{2010}
\maketitle

\label{firstpage}

\begin{abstract}
We present the results of the analysis of a sample of 17 low-luminosity ($L_{\rm X}~\lesssim~10^{42}\rm~erg/s$), radio loud AGNs in massive galaxies. The sample is extracted from the SDSS database and it spans uniformly a wide range in optical \OIII\ emission line and radio luminosity, but within a narrow redshift range ($0.05 < z < 0.11$) and a narrow super massive black hole mass range ($\sim 10^8 M_\odot$). For these sources we measured core X-ray emission with the \textit{Chandra X-ray telescope} and radio emission with the VLA. Our main goal is to establish which emission component, if any, can be regarded as the most reliable accretion/jet-power estimator at these regimes. In order to do so, we studied the correlation between emission line properties, radio luminosity, radio spectral slopes and X-ray luminosity, as well as more complex multi-variate relations involving black hole mass, such as the fundamental plane of black hole activity. We find that 15 out of 17 sources of our sample can be classified as Low-Excitation Galaxies (LEG) and their observed properties suggest X-ray and radio emission to originate from the jet basis. We also find that X-ray emission does not appear to be affected by nuclear obscuration and can be used as a reliable jet-power estimator. More generally, X-ray, radio and optical emission appear to be related, although no tight correlation is found. In accordance with a number of recent studies of this class of objects these findings may be explained by a lack of cold (molecular) gaseous structures in the innermost region of these massive galaxies.
\end{abstract}


\section{Introduction}

In this work we focus on three accretion/jet-power estimators among the different signatures of galactic nuclear activity arising at different wavelengths: optical line emission, nuclear X-rays and nuclear non-thermal radio emission. In the standard ``unified model'' for AGNs \citep{Urry1995}, optical narrow emission lines come from gas located several hundreds pc away from the central engine. These lines are excited by ionizing radiation produced in the innermost accretion flow and escaping along the polar axis of the obscuring torus that surrounds the black hole. Since this ionized gas is so distant from the central engine, the obscuring torus does not affect greatly its flux, thus narrow optical emission lines suffer only moderate amounts of dust obscuration due to the interstellar medium. This suggests that \OIII\ emission line luminosity can be a good estimator for the AGN accretion power. On the contrary, X-ray emission arises directly from the hot corona surrounding the accretion disk or from the base of a relativistic jet (for radio-loud objects). As such, it represents a more faithful estimator of the accretion power, although it could be heavily obscured by high column density through the dusty torus in objects where we look at the AGN from a close to edge-on sight-line. Finally, powerful non-thermal radio emission is the observational signature of the presence of a jet whose relativistic particles emit synchrotron radiation going through strong magnetic fields. Radio emitting jets are ubiquitous, particularly at low intrinsic powers \citep[see e.g.][and references therein]{Ho2008}, and it has been postulated that it also could be used to estimate the AGN kinetic and total power output \citep{Falcke1995, Best2006}.

As emerged in recent years \citep[see][and references therein]{Hardcastle2009}, radio loud objects can be divided into two main populations:
\begin{description}
 \item[\textit{high excitation galaxies (HEG):}] quasars and high-power narrow line radio galaxies that are likely to be the same population, seen at different orientation;
 \item[\textit{low excitation galaxies (LEG):}] active radio galaxies of low intrinsic power characterized by emission line spectra of low excitation, that probably form the parent population of the mostly lineless BL-Lac objects.
\end{description}
Recent studies \citep{Chiaberge2002a, Whysong2004, evans2006, Hardcastle2006, Hardcastle2009} have proposed that LEG lack any of the conventional apparatus of the AGNs (i.e. radiatively efficient accretion disk, X-ray corona and obscuring torus) and that their radio and X-ray nuclear emission can be explained as a result of the properties of the small-scale jet.

Significant progress has also been achieved in understanding how the radio properties of AGN relate to other parameters such as black hole mass and accretion rate \citep{Ho2002}. \cite{Merloni2003} (hereafter \citetalias{Merloni2003}) and \cite{Falcke2004a} considered heterogeneous black hole samples spanning a broad range in black hole mass and X-ray luminosity, and found a strong correlation between the radio luminosity $L_{\rm R}$ from the unresolved radio cores of AGNs, the black hole mass M and the 2-10 keV X-ray luminosity $L_{\rm X}$ of the form \citepalias{Merloni2003}:
\begin{equation}
 \log L_{\rm R} = 7.44 + 0.6 \log L_{\rm x} + 0.78 \log (M/M_{\odot}).
\end{equation}
These scalings with mass and luminosity can be well explained by the theory of synchrotron emitting compact relativistic jets and radiatively inefficient accretion flows \citep{Blandford1979, Heinz2003}. These works explicitly demonstrated that the black hole mass should be considered as a fundamental parameter in the determination of the observed relations between emission components in various bands.

In this paper we want to push forward these investigations on the relations between different accretion and power output estimators in super massive black hole (SMBH). In particular, we will test whether the loose empirical \OIII\ -- X-ray correlation found in type 1 AGN, as well as the fundamental plane of black hole activity, hold in an unbiased sample of type 2 AGN in the local universe. Compared to previous work on this subject \citep[see e.g.][hereafter \citetalias{Heckman2005}]{Heckman2005}, this study is unique because we select objects in a very narrow range of black hole
masses, thus disentangling the effects of mass and accretion rate on the observed \OIII~-~X-ray~-~radio correlations. 

The paper is organised as follows: in the next section we describe our sample, while in Section~\ref{sec:data_analisys} we lay out the most important calibration steps used for the X-ray, radio and optical data reduction. Then in Section~\ref{sec:comparison} we examine the
relation between different power estimators. Finally in Section~\ref{sec:discussion} we discuss their implications.

\section{Sample selection}
\label{sec:selection}
In order to answer some of the important questions posed in the previous section, we have performed a snapshot X-ray/radio survey of a new, well defined sample of low-luminosity AGN.
The sample used for our analysis was originally extracted from 2712 radio luminous type 2 AGN \citep{Best2005} contained in the 212000 galaxies of the second data release of the SDSS. From this sample we selected 17 objects with the following characteristics:

\begin{itemize}
 \item The black hole masses have been derived from velocity dispersion measurements of each galaxy \citep{Tremaine2002a, Heckman2004}. We require black hole masses to be $8 \leq \log M_{\rm BH} \leq 8.5$ (the range is comparable to the uncertainty in the estimate of $M_{\rm BH}$). This particular window was chosen to maximize the number of objects with $z \le 0.11$ that are radio loud.
 \item A second cut was made on the redshift: $0.05 \le z \leq 0.11$. The SDSS spectra are obtained using 3 arcsec diameter fibres and at larger distances, the AGN emission is significantly contaminated by starlight from the host galaxy.
 \item The objects were chosen to span a wide range in dust-corrected \OIII\ luminosity ($L_{\OIII} \sim 10^{5.5} L_\odot$ up to to $L_{\OIII} = 10^{9.2} L_\odot$), and the entire observed range in radio luminosity, $10^{38.5} {\rm erg\ s}^{-1} \le L_{\rm R} \le 10^{41.3} {\rm erg\ s}^{-1}$.
 \item In order to not be biased by cluster environment and to facilitate the X-ray data reduction, we chose objects that were at least 7 arcminutes away from the centre of galaxy clusters.
\end{itemize}
142 AGN satisfy all these conditions. Because of their intrinsic weakness, a full multi-wavelength survey of the entire sample was deemed too time-consuming. Instead, we chose to focus on a randomly selected sub-sample of 17 objects to cover the $L_{\OIII}$ -- $L_{\rm R}$ parameter space widely. The final choice of the sample size was made by trading
off the total observing time requested for maximizing the chances of source detection in the X-rays and ensuring that enough sources are included for statistical tests.

\begin{figure}
\includegraphics[width=0.5\textwidth]{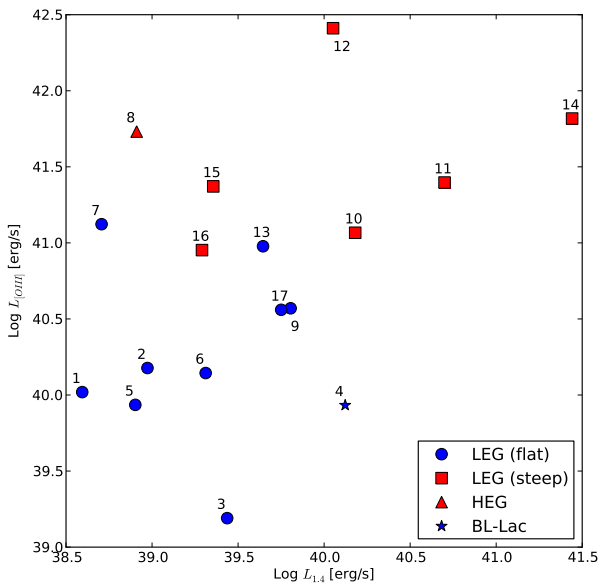}
\caption{Sources in our sample plotted in a radio (1.4 GHz) -- optical (\OIII) luminosity plane. Sources are selected in order to cover wide range of luminosity both in the radio and the optical. Points are coloured according to their spectral index in the radio band (red: steep-spectrum -- blue: flat-spectrum, see section~\ref{sec:radio} for details). Square markers are for LEG steep-spectrum objects, round markers are for LEG flat-spectrum objects while the triangle marker is the HEG and the star marker is the BL-Lac. For the classification criteria see section~\ref{sec:radio} and \ref{sec:optical}}
\label{fig:o3-radioFirst}
\end{figure}

Properties of the galaxies in the selected sample are listed in Table~\ref{tab:sample} while in Figure~\ref{fig:o3-radioFirst} we plot the sample on the \OIII\ -- radio plane using line luminosities from SDSS data release 7. At the time of the sample selection the last SDSS release available was the fourth. Subsequent analysis and better calibration of the SDSS spectroscopic data imply that some \OIII\ luminosity values differ slightly between the two releases due to different data reduction procedures and this is the reason why two objects (9 and 17) are quite near in the \OIII\ -- radio plane while, as stated above, we selected them to cover that space as uniformly as possible. Finally, we stress that in the sample selection process we did not exclude any source on the basis of the ionization state of the emission line complex, so that no bias for or against the HEG/LEG nature of the sources had been introduced. After our observations were taken, \cite{Plotkin2008}, using radio and optical data, have classified one of them as a BL-Lac with a strong boosting in the radio and X-ray emission. This is source number 4 in Table~\ref{tab:sample}, and is a clear outlier in most of the diagrams present throughout the paper. We decided to leave it in the sample for completeness of information.

Throughout this paper luminosity distances are calculated adopting $H_0 = 70\ {\rm km\ s}^{-1} {\rm Mpc}^{-1}$, $\Omega_m=0.3$ and $\Omega_\Lambda=0.7$.

\begin{table*}
\setcounter{table}{0}
\centering
\caption{The sample}
\label{tab:sample}
\begin{tabular}{lccccccccccc}
\hline
Object & ID & RA & DEC & z & $\log L_{\OIII}$ & $\log L_{\rm X}$ & \multicolumn{3}{c}{$\log L_{\rm R}$} & $\alpha_{\rm R}$ & $\log M_{\rm BH}$\\
 & & J2000 & J2000 & & & 2-10 keV & 1.4 GHz & 4.8 GHz & 8.4 GHz & &  \\
(1) & (2) & & & (3) &(4) & (5) & (6) & (7) & (8) & (9) & (10)\\
\hline
2MAS1601+46 & 1 & 16:01:09.07 & 46:23:17.8 & 0.08 & $40.0\pm0.13$ & $41.0^{+0.34}_{-0.84}$ & 38.6 & 39.2 & 39.4 & 0.12 & 8.0 \\
2MAS0837+53 & 2 & 08:37:17.99 & 53:15:16.8 & 0.06 & $40.2\pm0.08$ & $<41.5$ & 39.0 & 39.9 & 40.1 & -0.08 & 8.3 \\
2MAS1349+05 & 3 & 13:49:07.22 & 05:04:12.1 & 0.08 & $39.2\pm0.30$ & $41.0^{+0.38}_{-0.34}$ & 39.4 & 39.7 & 39.9 & 0.23 & 8.0 \\
CGCG043-05 & 4 & 12:53:47.01 & 03:26:30.4  & 0.07 & $39.9\pm0.07$ & $43.2^{+0.20}_{-0.17}$ & 40.1 & 40.6 & 40.8 & 0.17 & 8.5 \\
SDSS0141-09 & 5 & 01:41:16.35 & -08:35:39.3 & 0.05 & $39.9\pm0.04$ & $<41.0$ & 38.9 & 39.4 & 39.6 & 0.05 & 8.0 \\
2MAS1612+00 & 6 & 16:12:09.29 & 00:03:33.1 & 0.06 & $40.1\pm0.06$ & $41.3^{+0.18}_{-0.88}$ & 39.3 & 40.0 & 40.2 & 0.16 & 8.4 \\
SDSS2122-08 & 7 & 21:25:12.48 & -07:13:29.9  & 0.06 & $41.1\pm0.01$ & $42.3^{+0.11}_{-0.13}$ & 38.7 & 38.9 & 39.2 & -0.13 & 8.3 \\
2MAS1109+02 & 8 & 11:09:57.14 & 02:01:38.6 & 0.06 & $41.7\pm0.00$ & $42.6^{+0.67}_{-0.76}$ & 38.9 & 39.1 & 39.1 & 0.94 & 8.0 \\
2MAS0836+53 & 9 & 08:36:42.83 & 53:34:32.5  & 0.10 & $40.6\pm0.03$ & $41.7^{+0.20}_{-0.23}$ & 39.8 & 40.6 & 40.8 & 0.08 & 8.5 \\
2MAS1349+04 & 10 & 13:49:09.63 & 04:04:48.3 & 0.08 & $41.1\pm0.04$ & $40.4^{+0.70}_{-1.19}$ & 40.2 & 40.2 & 40.2 & 1.04 & 8.3 \\
2MAS0912+53 & 11 & 09:12:01.68 & 53:20:36.6 & 0.10 & $41.4\pm0.01$ & $41.2^{+0.32}_{-0.60}$ & 40.7 & 41.0 & 41.2 & 0.41 & 8.3 \\
2MAS0810+48 & 12 & 08:10:40.28 & 48:12:33.2 & 0.08 & $42.4\pm0.01$ & $41.1^{+0.72}_{-0.33}$ & 40.1 & 40.2 & 40.2 & 0.94 & 8.1 \\
2MAS1542+52 & 13 & 15:42:28.36 & 52:59:50.9 & 0.07 & $41.0\pm0.03$ & $42.0^{+0.37}_{-0.32}$ & 39.6 & 40.1 & 40.6 & -1.14 & 8.5 \\
4C52.370 & 14 & 16:02:46.39 & 52:43:58.4 & 0.11 & $41.8\pm0.03$ & $41.7^{+0.31}_{-0.37}$ & 41.4 & 41.5 & 41.5 & 0.94 & 8.5 \\
LCRS1010-02 & 15 & 10:12:39.87 & -01:06:22.9 & 0.10 & $41.4\pm0.03$ & $41.1^{+0.63}_{-0.39}$ & 39.4 & 39.9 & 40.1 & 0.32 & 8.2 \\
2MAS0101-00 & 16 & 01:01:01.11 & -00:24:44.4 & 0.10 & $41.0\pm0.02$ & $40.9^{+0.34}_{-1.00}$ & 39.3 & 39.7 & 39.8 & 0.72 & 8.4 \\
SDSS2305-10$^{\dag}$ & 17 & 23:08:17.29 & -09:46:22.5 & 0.10 & $40.6\pm0.07$ & $41.3^{+0.28}_{-0.25}$ & 39.8 & 40.1 & 40.4 & 0.10 & 8.3 \\
\hline
\end{tabular}
\vskip 0.3cm NOTE: Col.~(1): Name of the object. Col.~(2): Identification number. Col.~(3): Redshift. Col.~(4): \OIII\ luminosity $\log(L_{\OIII} {\rm[erg/s]})$ corrected for dust absorption. Col.~(5): X luminosity $\log(L_{\rm X} {\rm[erg/s]})$ (corrected for absorption). Col.~(6): Radio luminosity at 1.4 GHz $\log(L_{\rm 1.4} {\rm[erg/s]})$. Col.~(7): Radio luminosity at 4.8 GHz $\log(L_{\rm 4.8} {\rm[erg/s]})$ (mean 5 $\sigma$ error: 0.023 dex). Col.~(8): Radio luminosity at 8.4 GHz $\log(L_{\rm 8.4} {\rm[erg/s]})$ (mean 5 $\sigma$ error: 0.024 dex). Col.~(9): Radio spectral index, defined with $F \propto \nu^{-\alpha}$. Col~(10): SMBH mass $\log(M {\rm[M_{\odot}]})$. Object signed with a $\dag$ is resolved in the radio maps.
\end{table*}

\section{Data Analysis}
\label{sec:data_analisys}

Our sample of 17 objects has been observed with \textit{Chandra X-ray telescope} (see Section~\ref{sec:x-ray}) and with the VLA (see Section~\ref{sec:radio}). Information about their \OIII\ luminosities is taken from SDSS and is analysed as explained in Section~\ref{sec:optical}.

All errors due to inaccurate evaluation of distances are considered
negligible, while we account for an error of 0.25~dex in the black hole
mass \citep{Tremaine2002a}. Throughout the paper linear
fits have been performed using the Buckley-James method
\citep{BUCKLEY1979, Isobe1986} to account for upper limits. With this
regression algorithm, the most probable values of the upper limits are evaluated
through the Kaplan-Meier non-parametric distribution
\citep{Kaplan1958} and then used as observed values for a linear fit
that is performed using the orthogonal distance regression to take
into account errors both in the dependent and in the independent
variable. Correlation tests were performed using a generalized Kendall
correlation test \citep{Isobe1986} that accounts also for upper limits
and, if needed, for partial correlation \citep{Akritas1996}.

\subsection{X-ray}
\label{sec:x-ray}

All of the AGN were observed on the S3 chip of the Advanced CCD Imaging Spectrometer \citep[ACIS;][]{Garmire2003} aboard the \textit{Chandra X-ray Observatory}.  Data were taken in timed exposure mode with the standard frame time at the default location on
the S3 chip and telemetered to the ground in very faint mode.  Data reduction and point source extractions were completed using CIAO versions 4.2, and ACIS Extract version 2010-02-26 \citep[AE;][]{Broos2010}, respectively.
During extraction, the source position was adjusted based
on the mean position of the extracted counts to more accurately
calculate the point source photometry (except for the two sources with
upper limits). The position refinement was always well within \textit{Chandra}'s
absolute position uncertainty.

We used Sherpa version 4.2 to jointly fit the unbinned source and background spectrum of each source \citep{Freeman2001} using the C-statistic, which is similar to the \cite{Cash1979} statistic but with an approximate goodness-of-fit measure, and the {\small \sc moncar}\footnote{http://cxc.harvard.edu/sherpa4.2/ahelp/montecarlo.py.html} minimization algorithm. We used this statistical method for the fitting to avoid losing the little spectral information that we have for most of our spectra, which frequently have very few counts \citep[e.g.,][]{Nousek1989}. We fit each source with an absorbed power-law model ({\small \sc xszphabs $\times$ xspowerlaw}).  In all cases, the background exhibited the signs of hard particle background and was fit by two power laws and thermal component, which achieved a good fit. These ad hoc models are only used to constrain the background component of the spectra.

Since degeneracies in the fit parameters will frequently arise for very faint
sources, we followed a specialized scheme for these sources based on the number of
counts in the source extraction region. If we extracted less than 5 counts (0.5-8
keV) for a source, the power-law index and the hydrogen column density for the fit
were frozen to 1.7 and the Galactic value, respectively. If we extracted more than 5
but less than 26 counts (0.5-8 keV), we froze only the power-law index to 1.7 and
let the hydrogen column density float, although, in this case, it was always poorly
constrained. For all other sources with more than 26 counts (0.5-8 keV), we allowed
all fit parameters to float.

As can be seen in Table~\ref{tab:xray}, the AGN span a wide range of brightness.  We put upper
limits on the luminosities of two of the sources with only two counts each
(identified with number 2 and 5 in Table~\ref{tab:sample}).  On the other hand, there were also two
sources (IDs 4 and 7 in Table~\ref{tab:sample}) that had non-negligible pileup
fractions\footnote{http://cxc.harvard.edu/ciao/ahelp/acis\_pileup.html} (39.3\% and
4.6\% of the counts, respectively) and were therefore fit with the standard pileup
model.  Only two sources (number 8 and, to a lesser amount, 13) have significant
additional intrinsic absorption above that expected for the galactic foreground
\citep{Dickey1990}.  However, the determination of the column density for all of our sources is highly uncertain.

The unabsorbed X-ray luminosities listed in Table 1 were calculated from each of the
model fits and the uncertainties were calculated as follows.
The unabsorbed luminosity is evaluated at each point in a three-dimensional
grid 100 points on a side, each dimension corresponding to the number of
undetermined parameters in the source model (the absorption, power-law
photon index, and power-law normalization). Then the minimum and
maximum luminosity is selected within the confidence interval for the
appropriate change in statistic value \citep[$\Delta C = 3.53$ for one sigma
and three interesting parameters; e.g.,][]{avni1976}. The photon index is
constrained to be 0.7-2.7 and the column density must be at least the
galactic minimum in the direction of the source. In many cases, the uncertainties in
the luminosity estimated with this method are very different from those estimated
from the counts using Bayesian methods \citep{Kraft1991}.

Finally, we note that, since the \textit{Chandra} observations were carried out over
the span of one year (late 2006--early 2008), variability between the time of X-ray
observations and radio observations could affect some of our results.

\begin{table*}
\caption{X-ray data analysis values}
\label{tab:xray}
\begin{tabular}{lcccccccccc}
\hline
ID & ID & Obs. & Exposure & Source & Exp. bkg & Photon Index & $N_H$ [$10^{22} \mathrm{cm}^{-2}$] & Probability\\
 & \textit{Chandra} & date & time [s] & counts & counts & & & no source\\
(1) & (2) & & & (3) & (4) & (5) & (6) & (7)\\
\hline
1 & 8244 & 2008-01-16 & 11629 & 12 & 0.07 & 1.70 & $1.33_{-1.33}^{+14.3} \times 10^{-2}$ $^{\ddag}$ & $4.93 \times 10^{-23}$ \\
2 & 8245 & 2007-09-30 & 6938 &  2 & 0.06 & 1.70 & $3.51 \times 10^{-2}$ $^{\dag}$ & $1.52 \times 10^{-03}$ \\
3 & 8246 & 2007-03-15 & 10888 & 29 & 0.06 & $2.31_{-0.35}^{+1.85}$ & $1.99_{-1.99}^{+42.2} \times 10^{-2}$ $^{\ddag}$ & 0.00\\
4 & 8247 & 2007-03-04 & 8021 & 1044 & 0.10 & $2.53_{-0.37}^{+1.14}$ & $6.25_{-5.19}^{+11.5} \times 10^{-2}$ & 0.00\\
5 & 8248 & 2007-10-03 & 5109 & 2 & 0.03  & 1.70 & $3.04 \times 10^{-2}$ $^{\dag}$ & $3.47 \times 10^{-4}$\\
6 & 8249 & 2007-04-11 & 5950 & 22 & 0.04 & 1.70 & $7.42_{-7.42}^{+16.4} \times 10^{-2}$ $^{\ddag}$ & 0.00\\
7 & 8250 & 2007-06-29 & 6880 & 260 & 0.05 & $1.68_{-0.16}^{+0.21}$ & $6.00_{-6.00}^{+6.65} \times 10^{-2}$ & 0.00\\
8 & 8251 & 2007-04-10 & 6938 & 15 & 0.04 & 1.70 & $33.3_{-17.9}^{+14.7}$ & $2.87 \times 10^{-30}$\\
9 & 8252 & 2006-12-16 & 15933 & 101 & 0.13 & $2.23_{-0.19}^{+0.28}$ & $3.58_{-3.58}^{+6.67} \times 10^{-2}$ $^{\ddag}$ & 0.00\\
10 & 8253 & 2007-03-07 & 11005 & 3 & 0.06  & 1.70 & $1.97 \times 10^{-2}$ $^{\dag}$ & $3.09 \times 10^{-05}$\\
11 & 8254 & 2007-06-07 & 18782 & 21 & 0.15 & 1.70 & $1.78_{-1.78}^{+14.5} \times 10^{-2}$ $^{\ddag}$ & $3.32 \times 10^{-37}$\\
12 & 8255 & 2007-10-08 & 11008 & 12 & 0.07 & 1.70 & $1.67_{-1.67}^{+3.12} \times 10^{-1}$ & $7.36 \times 10^{-23}$\\
13 & 8256 & 2007-06-04 & 8912 & 20 & 0.05 & 1.70 & $4.25_{-1.26}^{+1.99}$ & $1.96 \times 10^{-44}$\\
14 & 8257 & 2007-06-03 & 19911 & 30 & 0.13 & $1.34_{-0.47}^{+1.10}$ & $1.14_{-1.14}^{+4.72} \times 10^{-1}$ & 0.00\\
15 & 8258 & 2007-01-13 & 16940 & 10 & 0.09 & 1.70 & $3.99_{-3.99}^{+7.53} \times 10^{-1}$ & $1.88 \times 10^{-17}$\\
16 & 8259 & 2008-02-09 & 16808 & 8 & 0.13  & 1.70 & $3.17_{-3.17}^{+41.6} \times 10^{-2}$ $^{\ddag}$ & $1.97 \times 10^{-12}$\\
17 & 8260 & 2007-10-20 & 17934 & 59 & 0.11 & $2.42_{-0.26}^{+0.27}$ & $3.09_{-3.09}^{+2.60} \times 10^{-2}$ $^{\ddag}$ & 0.00\\
\hline
\end{tabular}
\vskip 0.3cm 
NOTE: Col.~(1): Identification number. Col.~(2): Identification number used in \textit{Chandra} observations. Col.~(3): number of source counts (0.5 -- 8 keV) in the 90\% PSF. Col.~(4): expected background counts in source region (0.5 -- 8 keV). Col.~(5): is the best fits for the photon index and relative upper/lower limit increments. When the value is frozen to 1.7 (see text for details) the upper and lower limits are not calculable and therefore omitted. Col.~(6): is the column densities $N_H$ and relative upper/lower limit increments. $\dag$ indicates that $N_H$ is frozen to the galactic minimum. $\ddag$ indicates that the best fit happens to be at the galactic minimum. Col.~(7): ``prob\_no\_source'' value, defined in section 5.10.3 of the AE user manual (http://www.astro.psu.edu/xray/docs/TARA/ae\_users\_guide/). Sherpa's ``conf'' routine has been used to calculate the uncertainties in the photon index and column density (col. 5 and 6). Sherpa's ``conf'' method gives similar uncertainties to the grid method used in the calculation of the uncertainty in the luminosity, except that Sherpa's ``conf'' routine tends to underestimate the uncertainty relative to the grid method (e.g., on the order of a few tens of percent for the photon index).
\end{table*}

\subsection{Radio}
\label{sec:radio}

Radio data were taken in 2007 using the VLA in A configuration with a
total observation time of 10 hours. The calibration
procedure has been performed using CASA software (version 3.0.1) developed at
NRAO\footnote{http://casa.nrao.edu/}. At the time of the observations, 12 out of
26 antennas were already equipped with EVLA receivers. The presence of two
different types of receivers on the antennas caused different response from the
correlator whether a mixed baseline (VLA-EVLA) or a non-mixed one
(VLA-VLA and EVLA-EVLA) were processed. To solve the problem, a
baseline-based calibration was performed using the CASA task {\small \sc blcal}.
Finally, some cycles of self-calibration have been done on all sources.

We observed our sample in two different bands (4.8 and 8.4 GHz) and
from these flux values radio spectral indexes have been
extracted. Sources with a spectral index ($F \propto \nu^{-\alpha}$)
higher than $\alpha=0.3$ are classified as steep-spectrum, while other
sources are classified as flat-spectrum. Values from the FIRST survey
(1.4 GHz) were not taken into account during the spectral index
extraction process because source fluxes could have changed since FIRST
observations. From the spectral analysis we find that ten out of seventeen sources of our sample have a flat spectrum, while in one case (source 13), the spectrum is inverted.

To check the radio calibration procedure we computed the spectral index average values using the three different bands (1.4, 4.8 and 8.4 GHz) two by two. If the main driver for changes in the spectral shape between any two bands is time variability, then some of the sources would have an increased spectral index while some other a lowered one, with an average value of zero. This means that the three averaged
spectral index values should be similar. We find an averaged spectral index  $\alpha_{4.8-8.4}=-0.29\pm0.12$ between 4.8 and 8.4 GHz, $\alpha_{1.4-4.8}=-0.24\pm0.10$ between 1.4 and 4.8 GHz and $\alpha_{1.4-8.4}=-0.25\pm0.09$ between 1.4 and 8.4 GHz. They are all compatible, enforcing the idea that individual differences between spectral indices including the low-frequency FIRST observations are  mostly due to time variability.

The sample was selected to be unresolved at the FIRST resolution (5'')
and only one source (number 17) shows clear evidence of a
resolved structure in our new observations, that is a 1 arcsec long, jet-like feature in the west-north-west direction.
Finally radio fluxes were corrected to be all at the same rest-frame frequency using the derived
spectral indexes. Errors on radio fluxes have been set to 5 times the background RMS obtained analysing final maps.

\subsection{Optical}
\label{sec:optical}

\begin{figure*}
\begin{center}
\includegraphics[width=1\textwidth]{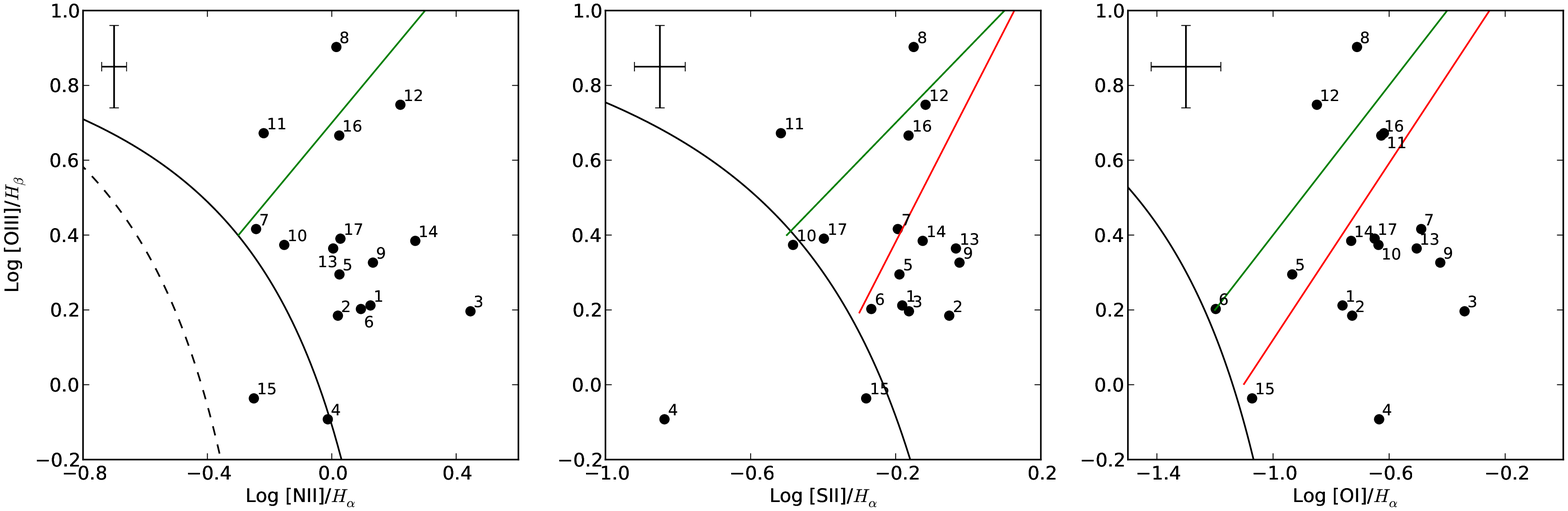}
\caption{Diagnostic diagrams for our 17 sources. Black solid lines \citep{Kewley2006} separate star forming galaxies, on the bottom left of each diagram, from AGNs. The region between dashed and solid lines \citep{Kewley2006} identify \textit{composite galaxies} with both star forming and AGN activity. Red lines \citep{Kewley2006} separate Seyfert galaxies (above) from LINERs (below). Green lines \citep{Buttiglione2009a} separate HEG (top) from LEG (bottom) sources. Average error bars are shown in the top left corner of each plot.}
\label{fig:diag}
\end{center}
\end{figure*}

We have re-analysed the available SDSS spectra for the sources in our
sample in order to assess the physical nature of the emission lines.
Optical spectroscopic information plays a major role in understanding
emission properties of central engine of radio galaxies. A first
classification using spectral lines was made by \cite{HeckmanT.M.1980}
and \cite{Baldwin1981}. They proposed to use optical lines ratios to
distinguish between HII regions ionized by star forming activities and
regions ionized by AGNs. Some years later \cite{Veilleux1987} revised
these definitions using only ratios of lines with a small separation
in wavelength, in order to reduce the reddening problem and the impact
of flux calibration errors. They introduced these lines combinations:
[OIII]$\rm \lambda5007/H\beta$, [NII]$\rm \lambda6583/H\alpha$,
[SII]$\rm \lambda\lambda6716,6731/H\alpha$ and
[OI]$\rm \lambda6364/H\alpha$. The separation between AGN and HII regions
was also calibrated by \cite{Kewley2001} and again by
\cite{Kewley2006} using $\sim 85\,000$ emission line galaxies from the
SDSS catalogue. Seyfert and LINERs occupy in these diagnostic diagrams
different places and the authors suggested that the observed dichotomy correspond to two
AGN subpopulations associated with different accretion modes.

Attempts to apply these diagnostic diagrams to
radio loud galaxies were made by \cite{LaingR.A.1994} on a
sub-sample of 3CR radio-galaxies. Following the suggestion by \cite{HineR.G.1979}
that FRII sources can be divided into two subclasses, they
separated their sample into high excitation galaxies (HEG, defined as
galaxies with $\rm \OIII/H\alpha > 0.2$ and equivalent width of $\OIII > 3\
\text{\AA}$) and low excitation galaxies (LEG). Following this idea, we have divided
our radio loud sample in HEG and LEG using the criteria described in
\cite{Buttiglione2009a}, where they take advantage from the combination
of the optical diagnostic planes. They define LEG all the sources with
$\rm\log(\OIII/H_\beta) - \left[ \log([NII]/H_\alpha) + \log([SII]/H_\alpha) + \log([OI]/H_\alpha) \right]/3 < 0.95$.

We calculated lines luminosities using the spectra extracted from the SDSS
data release 7 \citep{Abazajian2009a}. For our purposes we are interested both in the
\OIII\ observed line fluxes and in the dust corrected line fluxes. When dust attenuation
needs to be taken into account we measure $H\alpha$ and $H\beta$ line
intensities and use the Balmer decrement method. We assume an
intrinsic $H\alpha/H\beta$ of 2.87 \citep{OsterbrockDonaldE.1989}. The
attenuation correction we use is a double power law
\citep{Charlot2000} of the form: 
\begin{equation}
 \frac{\tau_\lambda}{\tau_\nu} = \left( 1-\mu \right) \left( \frac{\lambda}{5500 \text{\AA}} \right)^{-1.3} + \mu \left( \frac{\lambda}{5500 \text{\AA}} \right)^{-0.7}
\end{equation}
where $\tau_\nu$ is the total effective optical depth in the $V$
band. With this double power law we account for the attenuation due
the presence of discrete clouds random distributed (first term) and
the attenuation due to the ISM (second term), so $\mu$ is the fraction
of total $\tau_\nu$ caused by ambient interstellar medium. We set $\mu
= 0.3$ based on observed relations between UV continuum slope and the
$H\alpha$ to $H\beta$ line ratios.

We have generated emission line diagnostic \citep[BPT, after][]{Baldwin1981} diagrams to compare emission lines ratios
(\OIII$\lambda 5007 / \rm H\beta$, [OI]$\lambda 6003 / \rm H\alpha$,
[SII]$\lambda 6717,7631 / \rm H\alpha$ and [NII]$\lambda 6584 / \rm
H\alpha$) of our sample (see Figure~\ref{fig:diag}).

Following the previously explained classification from \cite{Buttiglione2009a} we can classify the majority of our sources as LEG and be confident only for source 8 to be classified as an HEG. Some sources (11 and 12) fall in an intermediate position since they do not satisfy requested line luminosity ratios in all diagnostic diagrams and they are then classified as LEG.
We must notice that with this division all sources optically classified
as LINERs are now classified as LEGs and source number 8, optically classified as a Seyfert, is an HEGs. The difference in the classifications is mainly for borderline objects (in our sample: 11, 12 and 16) that are conventionally classified as Seyfert galaxies but in \cite{Buttiglione2009a} classification are instead LEGs (see Figure~\ref{fig:diag}).
Contrary to what was found by \cite{Buttiglione2009a} we have some strong \OIII\ emitters ($L_{\OIII} > 10^{41.5}$) that do not have $\log\OIII / \rm H\beta \sim 1$, but lower values, e.g. source 12 and 14. Finally, object number 4 is an outlier in many plots. The reason is that this source, as classified by \cite{Plotkin2008} using radio and optical data, is a BL-Lac with a strong boosting in the radio and X-ray emission.

\section{Comparison among \OIII, X-ray and radio properties}
\label{sec:comparison}

\begin{table}
\caption{Partial correlation coefficients}
\label{tab:correlation}
\begin{tabular}{ccccc}
\hline
Indep. & Dep. & Test  & \multicolumn{2}{c}{Prob. of non correlation}\\
variable & variable & variable & observed & corrected\\
\hline
\multicolumn{5}{c}{All sample}\\
X-ray & Radio & [OIII] & 0.93 & 0.98 \\
Radio & [OIII] & X-ray & 0.29 & 0.06 \\
\OIII & X-ray & Radio & \textbf{0.01} & 0.09 \\
\hline
\multicolumn{5}{c}{LEG}\\
X-ray & Radio & [OIII] & 0.60 & 0.48 \\
Radio & [OIII] & X-ray & 0.12 & \textbf{$<$0.01} \\
\OIII & X-ray & Radio & 0.23 & 0.44  \\
\hline
\multicolumn{5}{c}{Radio steep-spectrum sources}\\
X-ray & Radio & [OIII] & 0.38 & 0.82 \\
Radio & [OIII] & X-ray & 0.22 & \textbf{0.02} \\
\OIII & X-ray & Radio & \textbf{$<$0.01} & 0.09 \\
\hline
\multicolumn{5}{c}{Radio flat-spectrum sources}\\
X-ray & Radio & [OIII] & 0.45 & 0.48 \\
Radio & [OIII] & X-ray & 0.87 & 0.88 \\
\OIII & X-ray & Radio & \textbf{0.05} & \textbf{0.05} \\
\hline
\end{tabular}
\vskip 0.3cm
NOTE: In boldface those values where the significance is $\geq 95\%$.\\
Object number 4 is not considered.
\end{table}

In this section we analyse the correlations and properties that
connect \OIII, X-ray and radio emissions from the sample of sources
described in Section~\ref{sec:selection}. As a preliminary step,
we have performed a partial correlation analysis on the sample (and
various sub-samples), both for the observed luminosities and for those
corrected for absorption/extinction, as discussed in the previous
section.
We adopt the \cite{Akritas1996} method to
control for censored data. Table~\ref{tab:correlation} shows a summary
of such a study, where the presence of a correlation between two
variables is tested accounting for the common dependence on the third
(test variable). In the next subsections
relations between different accretion power estimators are evaluated and
discussed in detail: in section~\ref{sec:accretion_estimators} we
present the analysis of the relationship between \OIII\ and X-ray
emission in our sample, while in
section~\ref{sec:radio_effect} we exploit the knowledge of radio
spectra to provide an interpretation of the X-ray and \OIII\ information. Finally,
in section~\ref{sec:plane} we discuss the X-ray -- radio correlation (or
lack of it) in light of previous studies on the fundamental plane of
active black holes.

\subsection{Accretion estimators: the relationship between \OIII\ and
  X-ray emission}
\label{sec:accretion_estimators}

\begin{figure*}
\includegraphics[width=\textwidth]{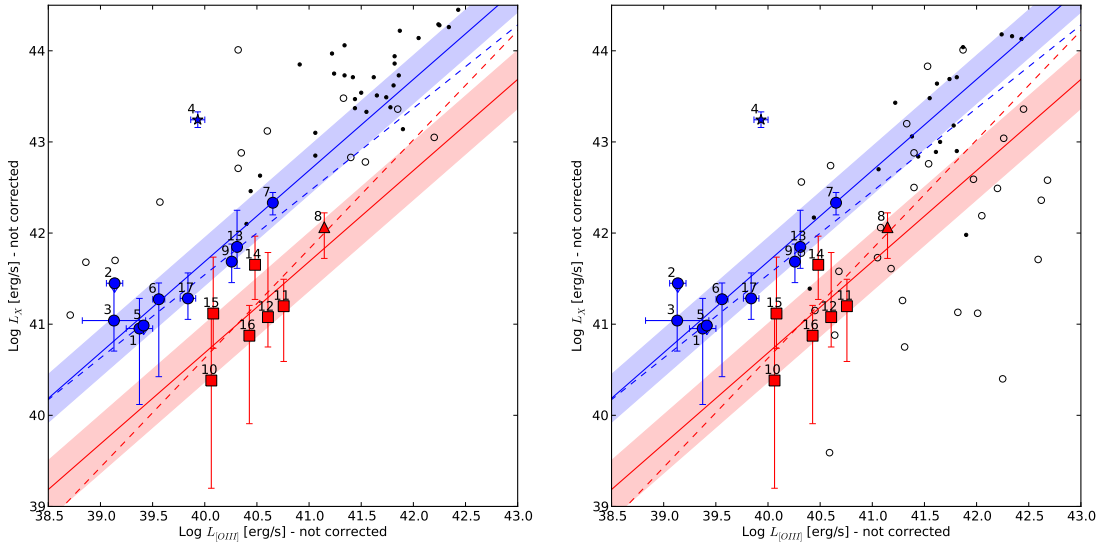}
\caption{Left panel: our sample on a X-ray [2-10 keV] -- optical (\OIII) plane (uncorrected for absorption). Symbols are as in Figure~\ref{fig:o3-radioFirst}. Black dots are Type 1 objects from \citetalias{Heckman2005} hard X-ray selected sample, while circles are Type 2. Red and blue solid lines are the mean ratio between $\log L_{\rm X}$ and $\log L_{\OIII}$ calculated using respectively steep and flat sources (coloured patches are at one sigma). Dashed lines are linear regressions (object number 4 is excluded). In the right panel the comparison is with the \OIII\ selected sample from \citetalias{Heckman2005}.}
\label{fig:o3-xray}
\end{figure*}

In \citetalias{Heckman2005} the authors emphasized the strong difference that arises if one considers
AGN samples of Seyfert galaxies selected by X-ray
luminosity or using the \OIII\ line luminosity. For a sample of
hard X-ray selected AGNs (flux limit in the 3-20 keV band of
$2.5\times10^{-11}\mathrm{\ erg\ s^{-1}\ cm^{-2}}$), they find a
mean $\log(L_{\rm X}/L_{\OIII})$ of 2.15 dex and $\sigma=0.51$
dex. Type 1 and Type 2 sources are separated in terms of
emission intensities, but do not present significant difference in the
mean ratio between X-ray and \OIII\ luminosities. In the left panel of
Figure~\ref{fig:o3-xray} the \citetalias{Heckman2005} X-ray selected dataset is shown together with our
sample, without applying any absorption/extinction corrections, for
ease of comparison (absorption-corrected correlations are
discussed in the following sections). Sources in our sample have a systematically lower X-ray to
\OIII\ ratio (see Table \ref{tab:ratio}), this is in agreement with
the different selection criteria used for the two samples. In the \citetalias{Heckman2005} X-ray selected sample a higher X-ray emission is indeed expected.

\begin{table}
\caption{$\log(L_{\rm X}/L_{\OIII})$ for uncorrected data}
\label{tab:ratio}
\begin{tabular}{ccc}
\hline
Sample & $\log(L_{\rm X}/L_{\OIII})$ & $\sigma$\\
 & [dex] & [dex] \\
\hline
\citetalias{Heckman2005} - X-ray selected & 2.15 & 0.51 \\
\citetalias{Heckman2005} - \OIII\ selected (Type 1) & 1.59 & 0.48 \\
\citetalias{Heckman2005} - \OIII\ selected (Type 2) & 0.57 & 1.06 \\
This work & 1.25 & 0.57 \\
\hline
\end{tabular}
\end{table}

In the right panel of Figure~\ref{fig:o3-xray} we compare instead our
dataset with the \OIII-selected sample of \citetalias{Heckman2005} (\OIII$\lambda5007$ fluxes
greater than $2.5 \times 10^{-13}\mathrm{erg\ cm^{-2}\ s^{-1}}$, corresponding to $L_{\OIII} \gtrsim 10^{40} \mathrm{erg\ s^{-1}}$). As
noted by \citetalias{Heckman2005}, Type 1 AGNs are detected at the expected rate,
while around 2/3 of the Type 2 AGNs would have been missed by the X-ray selection criteria. Their conclusion is
that the missing Type 2 AGNs are almost certainly heavily X-ray
absorbed. Our sample shows a different behaviour. Our objects in fact do not show a random
scatter in X-ray luminosity for a given \OIII\ line luminosity, rather
are quite aligned with the \citetalias{Heckman2005} \OIII-selected Type 1 sample. This can be
quantified in terms of the average luminosity ratios of the samples: in
\citetalias{Heckman2005} \OIII-selected Type 2 AGNs have an average $\log(L_{\rm
  X}/L_{\OIII})$ of 0.57 with a $\sigma$ of 1.06 dex. Our sample has
instead a higher X-ray to \OIII\ average ratio of 1.25 dex with a much
smaller $\sigma$ of 0.57 dex. Interestingly, this relation is much
more similar to what \citetalias{Heckman2005} find for the unobscured \OIII-selected AGNs
($\log(L_{\rm X}/L_{\OIII})=1.59$ dex, $\sigma = 0.48$ dex). 

If we exclude the peculiar BL-Lac object (number 4), the Kendall partial correlation
test shows an unexpected, although not strong, correlation between \OIII\
and X-ray emissions with a significance of 99\% (92\% if we keep the
BL-Lac). This correlation is not present in the Type 2 -- \OIII\
selected sample of \citetalias{Heckman2005}. A similar trend is instead present in the
X-ray selected sample, where X-ray absorption removes all the sources
in the lower right part of the X-ray -- \OIII\ plane, and in the Type 1
objects in both \citetalias{Heckman2005} samples, where obscuration is not important.

\begin{figure}
\includegraphics[width=0.5\textwidth]{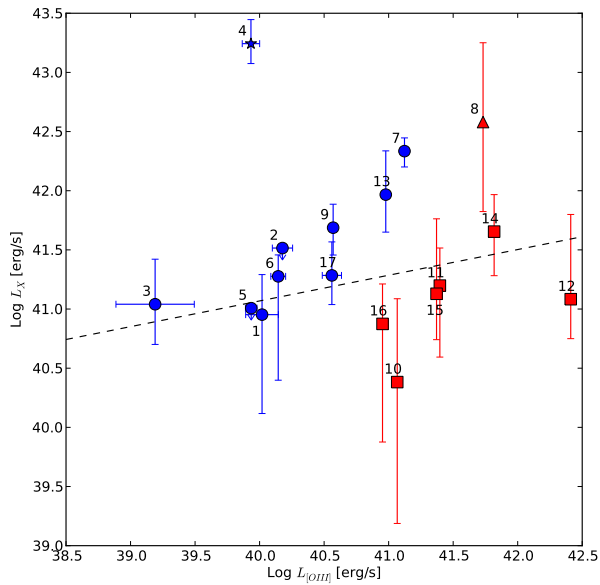}
\caption{The same relation shown in Figure~\ref{fig:o3-xray} after correction for absorption both in \OIII\ and X-ray. Black line is the linear regression (object 4 excluded, slope: $0.22\pm0.13$, no evident correlation present as stated in Table~\ref{tab:correlation}). Symbols are as in Figure~\ref{fig:o3-radioFirst}.}
\label{fig:o3-xray2}
\end{figure}

How can these difference be explained? 
The similarities between Seyfert 1 objects and our sample of low
luminosity AGNs in the \OIII-X-ray plane 
suggest a lack or a reduced incidence of molecular/dusty structures (like the
dusty torus present in Seyfert galaxies). The effect of such a structure would be to
absorb X-ray emission coming from the inner part of the AGN. This is
also confirmed by the X-ray spectral analysis, which does not find any
sign of absorption in any source of the sample (apart from number 8, that is
also the only HEG identified in the sample). The extinction-corrected \OIII\ -
X-ray plane, shown in Figure~\ref{fig:o3-xray2}, is indeed similar to
the uncorrected one, although due to the lack of high photon counts,
our absorption-corrected data have big error bars.

From this analysis we can conclude that, absorption being minimal, X-ray emission could be used as a good selection criteria for LEG sources at these low \OIII\ luminosities.

\subsection{The radio contribution}
\label{sec:radio_effect}

Our multi-wavelength dataset reveals another interesting fact.
We find that a part of the scatter in the X-ray -- \OIII\ relation can be
directly linked to the radio spectra of the sources. Splitting the sample into radio steep- ($\alpha \geq
0.3$) and flat-spectrum ($\alpha < 0.3$) sources, we find that the two sub-samples
 are still correlated in the \OIII\ -- X-ray plane (at a significance level $>$99\% for
flat-spectrum sources and of 95\% for steep-spectrum sources, see
Table~\ref{tab:correlation}) and their linear fits slopes are
compatible (flat: $0.91\pm0.09$ -- steep: $1.2\pm0.5$). But the
steep-spectrum sources are systematically ($\sim 1$ order of
magnitude) more \OIII\ luminous than flat-spectrum sources for a given
X-ray luminosity. Radio emission spectrum thus plays a role in
the location of the source in the $L_{\OIII}$ -- $L_{\rm X}$ plane, both
for the dust-uncorrected (Figure~\ref{fig:o3-xray}) and for the
dust-corrected one (Figure~\ref{fig:o3-xray2}).

One possibility is that our flat-spectrum sources occur in a gas-poorer
environment, consequently their jets undergo a small interaction with
the medium, so that, in the radio band, the very core of the AGN is the brightest part
of the object. The flat spectrum is then explained as superimposition
of many self-absorbed synchrotron spectra generated by the AGN core \citep{Blandford1979}.
These sources will also have a fainter \OIII\ emission because there is less gas
to ionize. This picture is reinforced by the fact that sources
with steep spectrum show on average more dust extinction. In these sources \OIII\ flux
is corrected for absorption (with the Balmer decrement method described in Section~\ref{sec:optical}) by a factor of $17\pm7$ on average, while flat-spectrum sources have an average correction of a factor $2.9\pm0.8$.

\begin{figure}
\includegraphics[width=0.5\textwidth]{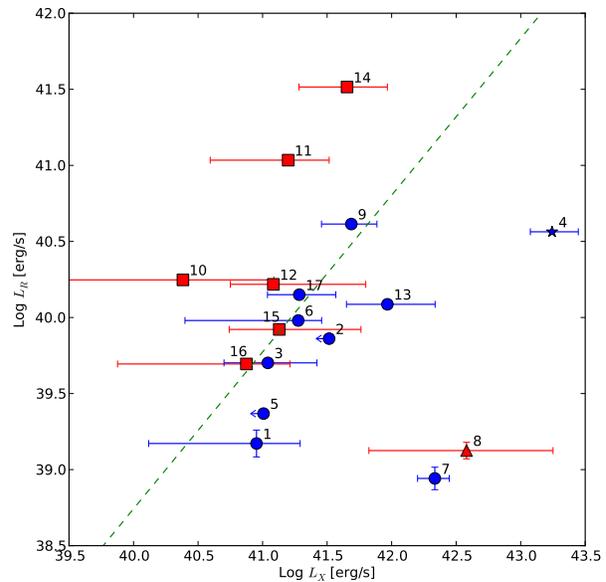}
\caption{X-ray -- radio (4 GHz) correlation. Green line is from a sample of Low Luminosity Radio Galaxy \citep{Panessa2007}. Symbols are as in Figure~\ref{fig:o3-radioFirst}.}
\label{fig:xray-radio}
\end{figure}

In the radio -- X-ray plane (Figure~\ref{fig:xray-radio}) there is no
evident correlation between the two variables and the separation
between steep- and flat-spectrum sources is not present any
longer. However, sources of our sample classified as LEG lie in the
same location (with the important exception of source 7) found by
\cite{Panessa2007} for a sample of Low Luminosity Radio Galaxy (green
line in Figure~\ref{fig:xray-radio}) extracted from the FR~I galaxies
of the 3C catalogue \citep{Balmaverde2006}. FR~I galaxies are indeed
almost always classified as LEG, and Seyfert galaxies are usually 2 -- 3
order of magnitude more powerful in X-ray for a given radio emission.

\subsection{The fundamental plane of active black holes}
\label{sec:plane}

\begin{figure*}
\includegraphics[width=\textwidth]{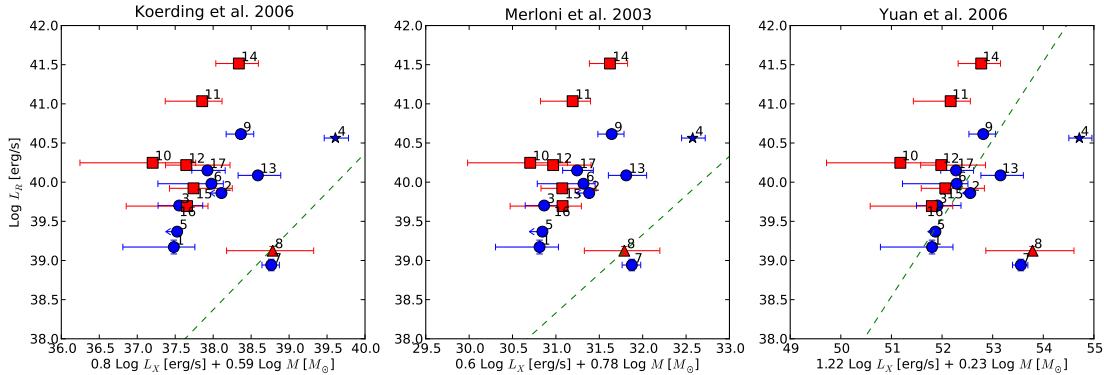}
\caption{Fundamental plane of active black hole relations. Green dashed lines shows the expected relation as found by \citet{Kording2006} (left), by \citetalias{Merloni2003} (centre) and by \citet{Yuan2009} (right). Symbols are as in Figure~\ref{fig:o3-radioFirst}.}
\label{fig:plane}
\end{figure*}

Almost all sources of our sample lie above the fundamental plane of
black hole activity described in \citet{Kording2006} (Figure~\ref{fig:plane}, left panel) and in \citetalias{Merloni2003} (Figure~\ref{fig:plane}, central panel).
These two planes were derived using AGNs selected with rather different criteria. For the first
using only AGNs where X-ray emission is believed to be jet-dominated while for the
second explicitly excluding jet-dominated systems.
An explanation of such behaviour in terms of underestimating
the X-ray extinction from the spectra and, therefore, a shift towards the
left side in Figure~\ref{fig:plane} is unlikely. Even taking into
account all difficulties in estimating the unabsorbed X-ray emission,
the vast majority of our sources (14 out of 17) would need to be so heavily
obscured as to bring their X-ray emission down by a factor
$\sim 100$. Such a large fraction of heavily obscured objects would be
very unusual, with current estimates of the incidence of Compton thick
AGN at these redshifts (albeit these are for somewhat higher luminosities) ranging between 20\% and 50\% \cite[see e.g.][]{Goulding2010}.

Also, the possibility of overestimating radio emission
due to boosting effect appears improbable. A boosting effect
should affect mainly sources with a flat spectrum, that is likely to be
connected with emission from AGNs core and jets, and would have
affected both radio and X-ray emissions, while what we find is an excess
of radio and a lack of X-ray emission. Furthermore, this would have
brought forth a much larger scatter in the fundamental plane relation and
the presence of some broadening in emission lines that are missing.

In fact, LEG sources in our sample describe a tighter relation (again with the exception of source 7)
than what expected from the fundamental plane of \citetalias{Merloni2003}. This relation has an intrinsic dispersion of $\sigma_\perp = 0.60$, while our LEG sample has a dispersion
that is a factor 1.7 smaller ($\sigma_\perp = 0.35$, that goes down to 
$\sigma_\perp = 0.26$ without source 7). The expected dispersion of the fundamental plane
of \citet{Kording2006} ($\sigma_\perp = 0.28$) is instead compatible with what we find.
It must also be noted that part of the dispersion in our data certainly arises from
the errors associated to (at least) our very weak X-ray luminosities
and consequent difficulties in estimating absorptions.

We find that neither \citetalias{Merloni2003} nor \citet{Kording2006} fundamental planes
are able to correctly recover the relation between radio and X-ray emissions for our sample of LLAGN.
This is indeed expected in the first case, where the majority of jet-dominated objects were
excluded from the derivation. In the \citet{Kording2006} version of the fundamental plane relationship
instead only jet-dominated system (low-hard state XRB, LLAGN and FRI radio galaxies) were exploited for derivation,
nevertheless our data do not follow that relation either.

Similar results were found by \cite{Hardcastle2009}, they tested the fundamental plane with a set of
powerful radio galaxies from the 3CRR catalogue whose jet and accretion X-ray components can
be disentangled using deep XMM spectroscopy. They found that, when
considering only the accretion-related X-ray component, their sources lie in the predicted
position on the fundamental plane of \citetalias{Merloni2003}. Using the jet related component
instead, the sources move to the same location where our sample of low
luminosity objects lie. This underline the care that needs to be taken before using
fundamental plane relations with objects that could be jet-dominated.

In \cite{Yuan2005} an interpretation of the fundamental plane in terms
of ADAF-jet model is given together with a prediction that the X-ray
emission should originate from jets rather than ADAFs
when the X-ray luminosity in the 2-10 keV band is lower than a
critical value $L_{\rm X,crit}$ \citep[equation 7 in][]{Yuan2005}. In
such a situation, the slope of the relation between radio and X-ray
luminosity is steeper and the normalization higher than that predicted
by \citetalias{Merloni2003}. \cite{Yuan2009} tested these ideas with a
sample of 22 low luminosity AGNs and found a relation very close to that defined by our sample
(left panel in Figure~\ref{fig:plane}) and close to the prediction of
\cite{Yuan2005} for X-ray emission produced at the base of the jet.
However, the critical X-ray luminosity predicted by \cite{Yuan2005},
for the object in our sample, should correspond to $\log L_{\rm X,crit}\approx
39.6$ erg/s, while all our objects have higher X-ray luminosities. 
Nevertheless, it should be noted that the \cite{Yuan2005} criterion
is to be intended in a statistical sense and not as a sharp division
between two populations, and the exact value of the critical
luminosity may depend on the specific incarnation of the ADAF
(viscosity, electron to ion heating efficiencies, etc.).
Even so, the location of our sources in the radio luminosity -
X-ray luminosity - SMBH mass space
enhance the interpretation that X-ray emission in our sample (at least
for LEG sources) is probably related to the jet synchrotron emission
rather than the Comptonization emission from the (radiatively inefficient)
accretion flow.

\section{Discussion}
\label{sec:discussion}

As we argued in Section~\ref{sec:accretion_estimators}, at least for LEG
sources, the absence or a reduced incidence of dusty/molecular gas may
explain the \OIII\ -- X-ray correlation and the unusual missing signs
of X-ray absorption in all LEG X-ray spectra (signs of absorption are
only found in the HEG source). Moreover, part of the scatter in
the observed \OIII-X-ray relations
is directly linked to the radio spectra of the sources, and we have argued in Section~\ref{sec:radio_effect} that
our flat-spectrum sources live in a gas poorer environment,
and consequently their jets undergo a small interaction with the ambient medium.

Independent evidence for an intrinsic difference in the amount of
cold/obscuring material in low luminosity AGN has been gathered
recently. \cite{Elitzur2009} showed that, in a nearly
complete sample of nearby AGN, the broad
line region disappears at low intrinsic luminosities, and interpreted
this fact within the
disk-wind scenario for the broad line region (BLR) and toroidal obscuration in AGN \citep{Elitzur2008}.
If the BLR constitutes the
inner, dust-free part of the cold molecular obscuring torus \citep[see e.g.][]{Netzer2008}, then a similar
behaviour should be expected also for other indicators of intrinsic
nuclear absorption. Indeed \cite{Tran2010}, using optical spectropolarimetry on a set of 3 low luminosity Type 2 AGNs, found no sign of broad line region nor obscuring torus. Furthermore, \cite{BurlonD.2010}
found for the first time tantalizing evidence that the fraction of
obscured AGN in a complete, hard-X-ray selected sample of nearby AGN,
declines at $L_{X}<10^{42}$ erg/s.

In Section~\ref{sec:plane} we pointed out a different behaviour
between LEG sources and those defining the \citetalias{Merloni2003}
fundamental plane, and this is explainable
with a different X-ray production mechanism between the population used to
construct the fundamental plane and our sample. In the fundamental
plane, X-ray emission is assumed to be generated by a (radiatively
inefficient) accretion flow
while, in LEG objects presented here, a much weaker (about 2 order of magnitude) X-ray
emission could be generated by the jet itself. The tight correlation
($\sigma_\perp = 0.35$) between radio and X-ray emission for LEG
objects enhances this conclusion, and is in agreement with the
analysis of \cite{Yuan2009}. Future studies on the relation
between radio and X-ray emissions and the object mass in a
bigger sample can unveil possible new relations similar to the
fundamental plane but tuned for low luminosity, jet-dominated
objects. Based on the results presented here, such a plane
will lie above the fundamental plane found in \citetalias{Merloni2003} and
will have a smaller scatter as a consequence of the tight relation
between radio and X-ray production in such kind of sources.

With respect to the analysis performed in \citetalias{Merloni2003}, we
remark here that the heterogeneity of their sample selection prevented
any uniform classification of the X-ray spectra in terms of the amount
of obscuration present. Based on the work presented here, it is
likely that a large part of the scatter in the original
fundamental plane relation could be due to the inclusion of both gas
poor and gas rich sources in the original analysis.

Taken together, these considerations open the possibility to use X-ray
emission as an unbiased estimator of source total power for
low luminosity AGN, provided robust indicators of a lack of cold
gas in the observed systems is available. \OIII\ emission, although useful in the
detecting and classification procedure, can be biased by the amount of
dust present in the AGN surroundings, as we can infer by different
position of steep- and flat-spectrum sources in
Figure~\ref{fig:o3-xray2}. 

\section{Conclusions}

In this paper we analysed a set of sources
extracted from the SDSS catalogue. For these sources we have
measured \textit{Chandra} X-ray flux, radio luminosity in
two bands from the VLA and optical spectral information
extracted from the SDSS database. Our sources span a wide range in optical
\OIII\ and radio luminosities but they have very similar SMBH masses ($\sim
10^8 M_\odot$) and are all located over a narrow redshift range ($0.05 < z < 0.11$).

Although our sample spans a wide range of
\OIII\ and radio luminosities, we found that 15 out of 17 sources are classified as LEG.
The sample shows a correlation on the uncorrected X-ray -- \OIII\ luminosity plane that is
not seen in samples of Type 2 AGN with higher \OIII\ luminosity \citepalias{Heckman2005}. An absent or
reduced obscuring torus can explain this relations and different
amount of gas in the AGN surroundings can account, to first order, for its scatter.
This last property is well
connected to the radio spectral index. This has important
implications for selecting criteria in upcoming
low-luminosity AGN surveys.
For LEG samples, the lack of X-ray obscuration means that X-ray
emission can be used as a good selection quantity with no
obscuration selection biases.
Nevertheless, radio core emission is still the
easiest way to identify this class of objects \citep{Ho2008},
although more information (i.e. from optical spectra) is
still required for object classification.

Summarizing, our conclusions:
\begin{itemize}
 \item At low luminosities ($L_{\rm X} <
   10^{42}$ erg/s) a selection criteria based on \OIII\ emission is not
   required in order to minimize the loss of of obscured
   sources. At low luminosity, X-ray and \OIII\ emissions
   are correlated and there are no evident signs of X-ray absorption.
 \item We found that all steep spectrum sources have around an
   order of magnitude less \OIII\ emission than flat-spectrum sources
   with similar X-ray luminosity. We argue that the amount of interstellar medium
   available for interaction in the proximity of the AGN is
   responsible for this distinction in the radio spectrum and for the
   difference in the production of the \OIII\ line.
 \item In the radio -- X-ray relation (and so in the fundamental
   plane of active black holes), LEG sources have a smaller dispersion and a reduced X-ray
   production compared to other types of AGN (i.e. Seyfert galaxies and
   quasars). This can be explained associating X-ray and radio
   production to the same physical mechanism, i.e. synchrotron
   emission from the base of the jet, in very good agreement with recent
   investigation \citep[see e.g.,][]{Hardcastle2009} although obtained
   from sample selected with different criteria.
\end{itemize}

\section*{Acknowledgements}
We are grateful to Angela Bongiorno for her useful help and
to ASTRON scientists, in particular John McKean, for their advices
on the radio data reduction.

\bibliographystyle{mn2e}
\bibliography{mn-jour,vla}
\bsp

\label{lastpage}

\end{document}